\documentclass[onecolumn,showpacs,superscriptaddress,nobibnotes,nofootinbib,12pt]{revtex4}

\usepackage{amsmath,amssymb}
\usepackage{graphicx}

\begin{document}

\title{On the exact solution of the accelerating string in $AdS_5$ space\footnote{
This work is supported in part by the US Department of Energy.}}
\author{Bo-Wen Xiao}
\email{bowen@phys.columbia.edu}
\affiliation{Department of Physics, Columbia University, New York, NY, 10027, USA}

\begin{abstract}
In this letter, the exact accelerating string solution of a heavy quark-antiquark pair is found in $AdS_5$ space. On the accelerating string, there is a particular scale which separates
the radiation and the heavy quark. This scale is explicitly shown to be an event horizon in the proper frame of the heavy quark. Furthermore, we find a new correspondence, which relates the horizon   in $AdS_5$ space on the gravity theory side to the Unruh temperature in Minkowski space on the field theory side of the AdS/CFT correspondence. $p_{\perp}$-broadening and $p_{L}$-broadening of the heavy quark due to radiation are computed using the AdS/CFT correspondence.
\end{abstract}

\date{\today}
\maketitle

\vfill

\newpage

\emph{Introduction.}---In Ref.~\cite{Dominguez:2008vd}, in order to describe bare quark energy loss in a finite size plasma, a brief description of the solution for an accelerating string in $AdS_5$ space was given, corresponding to a heavy quark-antiquark pair accelerated in opposite directions. In this paper, we will develop and describe in detail the theory of the accelerating string in $Ads_5$ spacetime\footnote{There is a similar numerical study of the accelerating string in Ref.~\cite{Herzog:2006gh} in the black three-brane metric of $AdS_5$ space. However, our focus in this letter is to study the exact accelerating string solution in vacuum $AdS_5$ spacetime. There is also a recent interesting study of the accelerating in Ref.~\cite{Chernicoff:2008sa} which considers a general time-dependent acceleration. The simplicity of our discussion comes about because we only consider constant acceleration in the vacuum for which we are able to find an exact analytic solution.}. The objective of this paper is to investigate the uniformly accelerating heavy quark-antiquark pair with a connecting string in $AdS_5$ spacetime. We find that there exists an event horizon on the string which separates the heavy quark and radiation during the acceleration. In other words, the upper part of the string is moving with the heavy quark, however, the lower part the string corresponds to radiated energy. Moreover, the event horizon is then shown to correspond to the well-known Unruh temperature\cite{Unruh:1976db} in a classical gravity calculation. In the end, the energy loss and $p_{\perp}$-broadening due to acceleration radiation are studied. 

\emph{The accelerating string solution}---We set up our accelerating string calculation as follows: a quark-antiquark pair is imbedded in a brane located at $u=u_{m}$, and a net
electric field $E_{f}$ is imposed in the brane which accelerates the quark and
antiquark at a constant acceleration in their own proper frame (An
additional small electric field $E_{f2}$ which balances the attracting force between the quark and
antiquark is also understood.).

The metric of the resulting vacuum $AdS_5$ space
can be written as 
\begin{eqnarray}
ds^{2} &=&R^{2}\left[ \frac{du^{2}}{u^{2}}-u^{2}dt^{2}+u^{2}\left(
dx^{2}+dy^{2}+dz^{2}\right) \right] \\
&=&\frac{R^{2}}{w^{2}}\left( dw^{2}-dt^{2}+dx^{2}+dy^{2}+dz^{2}\right) ,
\end{eqnarray}
where $R$ is the curvature radius of the $AdS_5$ space and $w=\frac{1}{u}$.
The dynamics of a classical string is characterized by the Nambo-Goto
action, 
\begin{equation}
S=-T_{0}\int d\tau d\sigma \sqrt{-\det g_{ab}}
\end{equation}
where $\left( \tau ,\sigma \right) $ are the string world-sheet coordinates,
and $-\det g_{ab}=-g$ is the determinant of the induced metric. $T_{0}$ is
the string tension. We define $X^{\mu }\left( \tau ,\sigma \right) $ as a
map from the string world-sheet to the five dimensional space time, and
introduce the following notation for derivatives: $\dot{X}^{\mu }=\partial
_{\tau }X^{\mu }$ and $X^{\prime \mu }=\partial _{\sigma }X^{\mu }$. When
one chooses a static gauge by setting $\left( \tau ,\sigma \right) =\left(
t,u\right) $, and defines $X^{\mu }=\left( t,u,x\left( t,u\right)
,0,0\right) $, it is straightforward to find that 
\begin{eqnarray}
-\det g_{ab} &=&\left( \dot{X}^{\mu }X_{\mu }^{\prime }\right) ^{2}-\left( 
\dot{X}^{\mu }\dot{X}_{\mu }\right) \left( X^{\prime \mu }X_{\mu }^{\prime
}\right)  \label{decayA} \\
&=&R^{4}\left( 1-\dot{x}^{2}+u^{4}x^{\prime 2}\right) .
\end{eqnarray}
Therefore, the equation of motion of the classical string reads
\begin{equation}
\frac{\partial }{\partial u}\left( \frac{u^{4}x^{\prime }}{\sqrt{-g}}\right)
-\frac{\partial }{\partial t}\left( \frac{\dot{x}}{\sqrt{-g}}\right)=0
\end{equation}

\begin{figure}[tbp]
\begin{center}
\includegraphics[width=9cm]{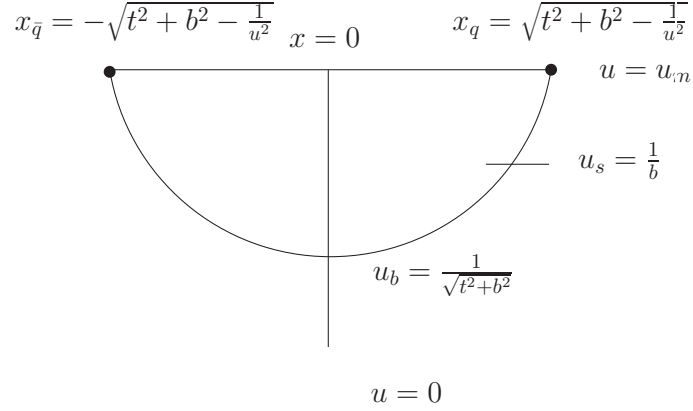}
\end{center}
\caption[*]{Illustrating the accelerating string.}
\label{accstring}
\end{figure}

In general, this equation is a non-linear differential equation which
involves two variables and two derivatives. Thus it is notoriously hard to
solve directly when $x\left( t,u\right) $ is a non-trivial function of $%
\left( t,u\right) $. Fortunately, we have been able to find the exact
solution which corresponds to the accelerating string. The solution reads, 
\begin{equation}
x=\pm \sqrt{t^{2}+b^{2}-\frac{1}{u^{2}}}  \label{solution}
\end{equation}
where the $+$ part represents the right moving part of the string and the $-$
part yields the left moving part of the string, together with the smooth
connection in the middle. The quark and antiquark pair are accelerating and
moving away from each other. The constant $b$ can be fixed by the boundary
condition. It is very easy to check that Eq.~(\ref{solution}) satisfies the
equation of motion by noting that $\sqrt{-g/R^4}=\frac{b}{\sqrt{t^{2}+b^{2}- 
\frac{1}{u^{2}}}}$.

Following Herzog et al \cite{Herzog:2006gh} , one can compute the canonical
momentum densities associated with the accelerating string, 
\begin{eqnarray}
\pi _{\mu }^{0} &=&-T_{0}\frac{\left( \dot{X}^{\nu }X_{\nu }^{\prime
}\right) X_{\mu }^{\prime }-\left( X^{\prime \nu }X_{\nu }^{\prime }\right) 
\dot{X}_{\mu }}{\sqrt{-g}}, \\
\pi _{\mu }^{1} &=&-T_{0}\frac{\left( \dot{X}^{\nu }X_{\nu }^{\prime
}\right) \dot{X}_{\mu }-\left( \dot{X}^{\nu }\dot{X}_{\nu }\right) X_{\mu
}^{\prime }}{\sqrt{-g}}.
\end{eqnarray}%
The energy density is given by $\pi _{t}^{0}$, 
\begin{equation}
\frac{dE}{du}=-\pi _{t}^{0}=\frac{T_{0}R^{4}}{\sqrt{-g}}\left(
1+u^{4}x^{\prime 2}\right) .
\end{equation}%
Thus the total energy of the right half of the string at time $t$ is,%
\begin{equation}
\int_{u_{b}}^{u_{m}}\frac{dE}{du}du=\frac{T_{0}R^{2}u_{m}}{b}\sqrt{%
t^{2}+b^{2}-\frac{1}{u_{m}^{2}}}.
\end{equation}
Moreover, the energy flow is given by $\pi _{t}^{1}$, 
\begin{equation}
\frac{dE}{dt}=\pi _{t}^{1}=\frac{T_{0}R^{4}}{\sqrt{-g}}u^{4}x^{\prime }\dot{x%
}.
\end{equation}%
Assuming that the quark carries a unite charge and $E_f$ being the external electric field as defined above, thus the force acting on the 
top of the string is just $qE_f=E_f$, and the energy being put into the system by the external force $E_f$ is $E_f (X(t)-X(0))$. Thus the net energy\footnote{The total energy being put into the system should be the sum of work done by $E_{f}$ and $E_{f2}$. However, the part from $E_{f2}$ balances the Coulomb potential between quark and antiquark as part of our setup in the beginning of the calculation. Thus only $E_{f}$ contributes to the non-Coulomb net energy increase.} being put into the right half string from $0$ to $t$ is,
\begin{eqnarray}
\left. \int_{0}^{t}\frac{dE}{dt}dt\right\vert _{u=u_{m}}&=&\frac{
T_{0}R^{2}u_{m}}{b}\left( \sqrt{t^{2}+b^{2}-\frac{1}{u_{m}^{2}}}-\sqrt{b^{2}-%
\frac{1}{u_{m}^{2}}}\right) \\
&=&\frac{
T_{0}R^{2}u_{m}}{b}\left(X(t)-X(0)\right),
\end{eqnarray}
with the second term in the bracket being the initial energy deposited in
the string. Also $b^{2}-\frac{1}{u_{m}^{2}}\geq 0$ is assumed for
consistency.  Therefore, from energy conservation, the energy increase of the system should be equal to $E_f (X(t)-X(0))$. One can easily fix the
constant $b$ by setting $E_{f}=\frac{T_{0}R^{2}u_{m}}{b}$, then, 
\begin{equation}
b=\frac{M}{E_{f}}=\frac{\sqrt{\lambda}u_{m}}{2\pi E_{f}},
\end{equation}
where $M=T_{0}R^{2}u_{m}$ is the mass of the heavy quark and $T_{0}R^2=\frac{%
\sqrt{\lambda}}{2\pi}$ according to the Ads/CFT correspondence\cite{Maldacena:1997re,Witten:1998qj,Gubser:1998bc}. It is now
very easy to see the physical interpretation of the constant $b$ as
the reciprocal of the constant acceleration $a$, i.e., $a=\frac{E_{f}}{M}=%
\frac{1}{b}$.

In addition, although $\frac{\partial x}{\partial t}=\frac{t}{\sqrt{
t^{2}+b^{2}-\frac{1}{u^{2}}}}$ exceeds $1$ when $u$ becomes smaller than $
1/b $, one can compute the speed which energy travels by the following,
\begin{equation}
v=\frac{\partial x}{\partial t}+\frac{\partial x}{\partial u}\frac{du}{dt}=%
\frac{t}{t^{2}+b^{2}}\sqrt{t^{2}+b^{2}-\frac{1}{u^{2}}},
\end{equation}%
and find that $v\leq 1$ at all times. In arriving at the above result, one
needs to look at the hypersurface where energy is constant (For example, one can focus on a lower segment of string with constant energy. One of the ends of this segment is taken as the bottom of the whole string and the other end can be taken to be somewhere $u<1/b$), 
\begin{equation}
E(u,t)=C \quad \Rightarrow \quad \frac{\partial E}{\partial t}+\frac{\partial E}{\partial u} \frac{du}{dt} =0
\end{equation}
Then, one can
obtain $\frac{du}{dt}=-\frac{\partial E}{\partial t}/\frac{\partial E}{%
\partial u}=- \frac{ut}{t^{2}+b^{2}}$ according to the energy flow along the string. Here $\frac{du}{dt}$ simply implies that the separation between the lower segment and the rest part of the string has to move downwards at the rate which we found above. Therefore, although some part of the string may travel 
with a speed beyond speed of light, the energy in the string can only travel at a speed smaller than speed of light. This indicates that the accelerating string solution is consistent with physical expectations. Finally, the Lorentz boost factor of
the string reads,%
\begin{equation}
\cosh \eta =\frac{1}{\sqrt{1-v^{2}}}=\frac{t^{2}+b^{2}}{\sqrt{\left(
t^{2}+b^{2}\right) b^{2}+\frac{t^{2}}{u^{2}}}},
\end{equation}%
and it reduces to $\frac{t}{b}$ in the large $t$ and $u$
limits.

\emph{The event horizon and the Unruh temperature}---In the following, we employ a transformation which transforms our system from $AdS_{5}$ to a generalized Rindler spacetime. To a uniformly accelerated observer, Minkowski spacetime becomes the so called Rinder spacetime. With properly chosen parameters, the heavy quark and the string look static in our generalized Rindler spacetime. In other words, we choose to transform to the proper frame of the accelerating string. This frame is a accelerating
frame with a constant acceleration $a$. The transform reads, 
\begin{eqnarray}
x &=&\sqrt{b^{2}-r^{2}}\exp \left( \frac{\alpha }{b}\right) \cosh \frac{\tau 
}{b}, \notag \\ 
t &=&\sqrt{b^{2}-r^{2}}\exp \left( \frac{\alpha }{b}\right) \sinh \frac{\tau 
}{b},  \notag \\
w &=&r\exp \left( \frac{\alpha }{b}\right) .  \label{mapping}
\end{eqnarray}
The new coordinates have ranges $-\infty < \alpha \text{,} \tau <\infty$ and $0<r< b$. It only covers the wedge $x>|t|$ for fixed $w$. Generally speaking, this transformation can cover the full $AdS_{5}$ space by including regions as in Eq.~(\ref{mapping}) but with different signs. Putting this transformation into Eq.~(\ref{solution}), one finds 
\begin{equation}
\alpha =0,
\end{equation}
which is now our equation of motion in the accelerating frame. This mapping
only maps the upper part of the string ($0<w<b$ or $1/b<u<\infty $) into the
proper frame of the accelerating quark. Substituting the variables into $%
\alpha $, $\tau $ and $r$ in the metric, one finds 
\begin{equation}
ds^{2}=\frac{R^{2}}{r^{2}}\left[ \frac{dr^{2}}{1-r^{2}/b^{2}}-\left(
1-r^{2}/b^{2}\right) d\tau ^{2}+d\alpha ^{2}+\left( dy^{2}+dz^{2}\right)
\exp \left( -2\alpha /b\right) \right]. \label{metric1}
\end{equation}
This metric contains an event horizon at $r=b$, which separates the string into 
two parts. There is no causal connection between these two parts of the string. An observer at $r=0$ can only see the part of the string from $r=0$ to $r=b$.  This result implies that our above transform is self-consistent. 

Furthermore, one can compute the well-known Hawking temperature\cite{Hawking:1974sw} which is determined by the
behavior of the metric near the horizon
\begin{equation}
T=-\frac{h^{\prime }\left( b\right) }{4\pi }=\frac{1}{2\pi b}=\frac{a}{2\pi }, \label{unruh}
\end{equation}
where $h\left( r\right) =1-r^{2}/b^{2}$. This temperature agrees
with the Unruh temperature, which was found in the quantum free scalar fields calculation\cite{Unruh:1976db} in an accelerating frame and now is believed to be true in more general circumstances. Physically, the Unruh effect means that the Minkowski vacuum looks like a state containing many particles in thermal equilibrium with a temperature given by Eq.~(\ref{unruh}) in Rindler spacetime. In our generalized Rindler spacetime, we obtain a static string in a thermal bath in contrast to the accelerating string in the original zero temperature $AdS_5$ spacetime. Moreover, this implies that the Unruh temperature, which is a temperature obtained by quantizing the Minkowski vacuum in Rindler spacetime on the field theory side, has a corresponding event horizon in the fifth dimension in $AdS_{5}$ spacetime on the gravity side. 

\emph{Energy loss due to radiation}---We have explicitly shown that there is a scale $u_{s}=\frac{1}{b}$
separating the soft part(the lower part) of the string from the hard part of
the string(the upper part). The upper part, which moves together with the
heavy quark, corresponds to the co-moving hard partons in the heavy quark
wave function; The lower part ($u<u_{s}$) of the string, which lies far
behind the heavy quark, is emitted radiation, and it is no longer part of
the heavy quark.

Therefore, the radiated energy at time $t$\footnote{In arriving such conclusion, there is an underlying assumption that the constant acceleration should be maintained all the time. Otherwise, the event horizon can not be established between the radiation and the heavy quarks.} is 
\begin{equation}
E_{\text{radiation}}=\int_{u_{b}}^{u_{s}}\frac{dE}{du}=\frac{\sqrt{\lambda }%
}{2\pi }\frac{t }{b^{2}}=\frac{\sqrt{\lambda }%
}{2\pi }a^2 t, \label{re}
\end{equation}
thus the radiation power reads, 
\begin{equation}
P=\frac{dE_{\text{radiation}}}{dt}=\frac{\sqrt{\lambda }}{2\pi }\frac{%
E_{f}^{2}}{M^{2}}.
\end{equation}
This is very similar to the answer for a classical charge particle accelerating in a constant electric field\cite{jackson}.

\emph{$p_{\perp}$-broadening and $p_{L}$-broadening}---Since the radiation induced by the acceleration is also stochastic, one expects to have similar $p_{T}$ and $p_{L}$ broadening as in Refs.~\cite{Gubser:2006nz,CasalderreySolana:2006rq,CasalderreySolana:2007qw}(For a review, see Ref.~\cite{CasalderreySolana:2007pr}.). Here in this section, we evaluate the $p_{\perp}$-broadening and $p_{L}$-broadening in a more sophisticated manner following Refs.~\cite{Gubser:2006nz,CasalderreySolana:2006rq,CasalderreySolana:2007qw}. In order to characterize the stochastic feature of the radiation, we use the Langevin equation\footnote{We assume to work in a frame that the heavy quark has no initial velocity. Thus the drag force is now absent in the equation of motion.}
\begin{equation}
\frac{dp_{i}}{dt}= \mathcal{F}_{i}
\end{equation}
to model the brownian motion of the heavy quark. Here $\mathcal{F}_{i}$ is the effective random force acting on the quark, with the correlation being 
\begin{equation}
\langle\mathcal{F}_{i}\left(t\right)\mathcal{F}_{j}\left(t^{\prime}\right)\rangle=\kappa \delta_{ij}\delta \left(t-t^{\prime}\right),
\end{equation}
in which $\kappa$ characterizes the strength of the random noise as well as the mean square momentum transfer during the radiation. Then it is straightforward to find that,
\begin{eqnarray}
\frac{dp_{T}^2}{dt}= 2 \kappa, \,  \, \frac{dp_{L}^2}{dt}=  \kappa, 
\end{eqnarray}
where the $2$ comes from the sum over $i=1,2$. 

To compute $p_{\perp}$-broadening and $p_{L}$-broadening, we need to transform to the generalized Rindler spacetime shown above, and do another transformation by
setting $u^{\prime}=\frac{r^{2}}{b^{2}}$. Then the metric becomes, 
\begin{equation}
ds^{2}=\frac{R^{2}}{b^{2}u^{\prime}}\left[ -f\left(u^{\prime}\right) d\tau ^{2}+d\alpha
^{2}+\left( dy^{2}+dz^{2}\right) \exp \left( -2\alpha /b\right) \right] +%
\frac{R^{2}}{4u^{\prime 2}}\frac{du^{\prime 2}}{f\left(u^{\prime}\right)}, 
\end{equation}
where $f\left(u\right)=1-u^{\prime} $. Let $(t,\sigma )=(\tau ,u^{\prime})$ be the new worldsheet coordinates in the proper
frame of the heavy quark, and $X^{\mu }=(\tau ,u^{\prime},\delta (\tau ,u^{\prime}),0,0)$. It
is very straightforward to discover that fluctuations in $\alpha ,y,z$
directions are essentially symmetric along the string trajectory $\alpha =0$%
. Therefore, one gets the same equation of motions for both transverse and
longitudinal fluctuations. The action which includes the fluctuations on the
string reads,
\begin{equation}
S=-T_{0}R^{2}\int d\tau du^{\prime}\frac{1}{2bu^{\prime 3/2}}\sqrt{1+\frac{4u^{\prime}f\left(
u^{\prime}\right) }{b^{2}}\delta ^{\prime 2}-\frac{1}{f\left( u^{\prime}\right) }\dot{\delta}%
^{2}}.
\end{equation}%
We here rescale all the variables to
dimensionless ones as follows: $\frac{\tau }{b}\Rightarrow \overline{\tau }$
and $\frac{\delta }{b}\Rightarrow \overline{\delta }$. Thus the action
becomes,%
\begin{eqnarray}
S &=&-T_{0}R^{2}\int d\overline{\tau }du^{\prime}\frac{1}{2u^{\prime 3/2}}\sqrt{1+4u^{\prime}f
\overline{\delta }^{\prime 2}-\frac{1}{f}\dot{\overline{\delta }}^{2}} \\
&\simeq &-T_{0}R^{2}\int d\overline{\tau }du^{\prime}\frac{1}{2u^{\prime 3/2}}\left( 1+2u^{\prime}f
\overline{\delta }^{\prime 2}-\frac{1}{2f}\dot{\overline{\delta }}
^{2}\right) .
\end{eqnarray}
when $\overline{\delta }\ll 1$. Therefore, the equation of motion for the
fluctuation on the string is 
\begin{equation}
\partial _{u^{\prime}}\left( 2u^{\prime}f\frac{\overline{\delta }^{\prime }}{u^{\prime3/2}}\right)
-\partial _{\overline{\tau }}\left( \frac{\dot{\overline{\delta }}}{2fu^{\prime 3/2}%
}\right) =0.
\end{equation}
Performing a Fourier transform, 
\begin{equation}
\overline{\delta }\left( u^{\prime},\overline{\tau }\right) =\int \frac{d\omega }{%
2\pi }\exp \left( -i\omega \overline{\tau }\right) \delta \left( \omega
\right) Y_{\omega }\left( u^{\prime}\right) ,
\end{equation}%
where $\delta \left( \omega \right) $ is defined as the Fourier transform of
fluctuation on the boundary. Thus, $Y_{\omega }\left( u^{\prime}=0\right) =1$. It is
straightforward to transform the equation of motion to 
\begin{equation}
Y_{\omega }^{\prime \prime }\left( u^{\prime}\right) -\frac{1+u^{\prime}}{2u^{\prime}f\left( u\right) }%
Y_{\omega }^{\prime }\left( u^{\prime}\right) +\frac{\omega ^{2}}{4u^{\prime}f^{2}\left(
u^{\prime}\right) }Y_{\omega }\left( u^{\prime}\right) =0.
\end{equation}
Near $u^{\prime}=1$ we find two independent solutions, $Y_{\omega }\left( u^{\prime}\right)
=\left( 1-u^{\prime}\right) ^{\pm i\omega /2}F_{\omega }\left( u^{\prime}\right) $. Requiring
the in-falling boundary condition, one chooses the solution $Y_{\omega
}^{-}\left( u^{\prime}\right) =\left( 1-u^{\prime}\right) ^{-i\omega /2}F_{\omega }^{-}\left(
u^{\prime}\right) $ falling into the black hole. The resulting differential equation
for $F_{\omega }^{-}\left( u^{\prime}\right) $ reads, 
\begin{equation}
F_{\omega }^{-\prime \prime }\left( u^{\prime}\right) +\left[ \frac{i\omega }{1-u^{\prime}}-
\frac{1+u^{\prime}}{2u^{\prime} \left( 1-u^{\prime}\right) }\right] F_{\omega }^{-\prime }\left(
u^{\prime}\right) +\left[ -\frac{i\omega }{4u^{\prime}\left( 1-u^{\prime}\right) }+\frac{\omega ^{2}}{
4u^{\prime}\left( 1-u^{\prime}\right) }\right] F_{\omega }^{-}\left( u^{\prime}\right) =0.
\end{equation}
This equation can be solved perturbatively in $\omega $. We find $F_{\omega }^{-}\left( u^{\prime}\right) =1+i\omega g\left( u^{\prime}\right) +o\left( \omega
^{2}\right),$ where $g\left( u^{\prime}\right) $ satisfies
\begin{equation}
g^{\prime \prime }\left( u^{\prime}\right) -\frac{1+u^{\prime}}{2u^{\prime}\left( 1-u^{\prime}\right) }g^{\prime
}\left( u\right) -\frac{1}{4u^{\prime}\left( 1-u^{\prime}\right) }=0.
\end{equation}
Requiring that $\left. g\left( u^{\prime}\right) \right\vert _{u^{\prime}=0}=0$ and $g\left(
u^{\prime}\right) $ is not singular at the horizon $u=1$, one reaches $g\left( u^{\prime}\right) =-\sqrt{u^{\prime}}+\ln \left( 1+\sqrt{u^{\prime}}\right).$ Therefore, we find 
\begin{equation}
Y_{\omega }\left( u^{\prime}\right) =\left( 1-u^{\prime}\right) ^{-i\omega /2}\left\{
1+i\omega \left[ -\sqrt{u^{\prime}}+\ln \left( 1+\sqrt{u^{\prime}}\right) \right] +o\left(
\omega ^{2}\right) \right\} .
\end{equation}%
Following Son et al\cite{Son:2007vk,Herzog:2002pc,Son:2002sd}, one can compute the retarded Green's function from the kinetic part of the action by using the Ads/CFT correspondence
\begin{eqnarray}
G_{R}\left( \omega \right)  &=&-2T_{0}R^{2}\frac{1}{b^{2}}\left. \frac{1}{%
u^{\prime 1/2}}f\left( u^{\prime}\right) Y_{-\omega }\left( u^{\prime}\right) \partial _{u^{\prime}}Y_{\omega
}\left( u^{\prime}\right) \right\vert _{u^{\prime}\rightarrow 0}, \\
&=&-2T_{0}R^{2}\frac{1}{b^{2}}\left( \frac{i\omega }{2}+o\left( \omega
^{2}\right) \right) ,
\end{eqnarray}%
where the retarded Green's function is defined as $iG_{R}\left(t\right)=\theta \left(t\right)\langle [\mathcal{F}\left(t\right),\mathcal{F}\left(0\right)] \rangle$. We have re-inserted units and restored the dimensions of $\delta \left( u^{\prime},t\right) $. Moreover, one can relate the Schwinger-Keldysh propagators and the retarded Green's function in the following way\cite{Herzog:2002pc}\footnote{Following Ref.~\cite{Herzog:2002pc}, one can start with the metric in Eq.~(\ref{metric1}) and replace the temperature by the Unruh temperature, and then derive Eq.~(\ref{kubo}) by examining the Kruskal diagram.}, 
\begin{eqnarray}
G_{11}\left( \omega \right) &=& \textrm{Re}G_{R}\left( \omega \right) + i \coth\frac{\omega}{2T} \textrm{Im}G_{R}\left( \omega \right), \\
G_{12}\left( \omega \right) &=&  G_{21}\left( \omega \right) =\frac{2i e^{-\omega/2T}}{1-e^{-\omega/T}} \textrm{Im}G_{R}\left( \omega \right),\\
G_{22}\left( \omega \right) &=& -\textrm{Re}G_{R}\left( \omega \right) + i \coth\frac{\omega}{2T} \textrm{Im}G_{R}\left( \omega \right).
\end{eqnarray}
Then, as in Refs.~\cite{Gubser:2006nz,CasalderreySolana:2006rq,CasalderreySolana:2007qw}, one finds 
\begin{eqnarray}
\kappa  &=&\lim_{\omega \rightarrow 0}\frac{1}{4} \left[
iG_{11}\left( \omega \right)+iG_{12}\left( \omega \right)+iG_{21}\left( \omega \right)+iG_{22}\left( \omega \right) \right] , \\
&=&\lim_{\omega \rightarrow 0}\frac{-2T}{\omega }\textrm{Im}
G_{R}\left( \omega \right).\label{kubo}
\end{eqnarray}
Eq.~(\ref{kubo}) is also known as Kubo's formula, and it is derived using the fluctuation-dissipation theorem. This formula is only true for systems in thermal equilibrium. Since the string becomes static and inhabits in a thermal bath in the generalized Rinder space, the temperature in Eq.~(\ref{kubo}) should be $\frac{a}{2\pi}$ according to the correspondence we found above. Thus,
\begin{eqnarray}
\kappa =\lim_{\omega \rightarrow 0}\frac{-2T}{\omega }\textrm{Im}
G_{R}\left( \omega \right) =\frac{\sqrt{\lambda }}{2\pi ^{2}}\frac{1}{b^{3}}=\frac{\sqrt{\lambda }}{%
2\pi ^{2}}a^{3}.
\end{eqnarray}%
In the last step we have inserted the relation $T_{0}R^{2}=\frac{\sqrt{%
\lambda }}{2\pi }$, and $T=\frac{1}{2\pi b}$. 

Finally, one can compute the $p_{T}$ and $p_{L}$ broadening due to radiation
in the instantaneous co-moving frame of the accelerating quark,
\begin{eqnarray}
\frac{d\widetilde{p}_{T}^{2}\left( \tau \right) }{d\tau } =2\kappa =\frac{%
\sqrt{\lambda }}{\pi ^{2}}a^{3}, \, \,
\frac{d\widetilde{p}_{L}^{2}\left( \tau \right) }{d\tau } =\kappa =\frac{%
\sqrt{\lambda }}{2\pi ^{2}}a^{3}.
\end{eqnarray}%
In the lab-frame, the $p_{T}$ and $p_{L}$ broadening of heavy quark read 
\begin{eqnarray}
\frac{dp_{T}^{2}\left( t\right) }{dt} =\frac{\sqrt{\lambda }}{\pi ^{2}}%
\frac{a^{3}}{\sqrt{a^{2}t^{2}+1}}, \,\,
\frac{dp_{L}^{2}\left( t\right) }{dt} =\frac{\sqrt{\lambda }}{2\pi ^{2}}%
a^{3}\sqrt{a^{2}t^{2}+1}. \label{ptff}
\end{eqnarray}%
In arriving above result, we have used the Lorentz boost factor $\gamma =\frac{\sqrt{t^{2}+b^{2}}
}{b}$. 

By using the partonic picture of Ref.~\cite{Dominguez:2008vd}, we could also estimate the $p_{T}$ and $p_{L}$ broadening due to radiation. This analysis may help to interpret the result that we found in Eq.~(\ref{ptff}). At large time limit, one finds 
\begin{equation}
\frac{dp_{T}^{2}}{dt}\propto \frac{\sqrt{\lambda }}{2\pi }\frac{u_{s}^{2}}{t}%
\sim \frac{\sqrt{\lambda }}{2\pi }\frac{1}{b^{2}t}, \label{pt1}
\end{equation}%
where $\sqrt{\lambda }$ could be interpreted as probability of emitting partons as part of the radiation.\footnote{$\sqrt{\lambda }$ is analogous to the factor $\alpha_s N_c$ which essentially gives probability of emitting a gluon from a quark in perturabative QCD. Here we would like to interpret the effective coupling $\sqrt{\lambda }$ as the probability of emitting a gluon from a heavy quark in strongly coupled SYM theory.} $u_{s}^{2}$ may be the typical momentum square of the emitted partons, and $t$ is just the time scale of the system. Similarly, one finds 
\begin{equation}
\frac{dp_{L}^{2}}{dt}\propto \frac{\sqrt{\lambda }}{2\pi }\frac{\omega
_{s}^{2}}{t}\sim \frac{\sqrt{\lambda }}{2\pi }\frac{t}{b^{4}}, \label{pl1}
\end{equation}%
with $\omega _{s}\sim \frac{1}{\Delta x}\simeq u_{s}^{2}t$ being the typical
energy of the emitted partons and $\Delta x$ being the longitudinal
separation between the quark and the string at $u=u_{s}$. Moreover, after
identifying $u_s$ with $Q_{s}$, one discovers that the coherence time $t=%
\frac{\omega_s}{u_{s}^2}$ in this accelerating string scenario coincides with
the one in QCD. 

\emph{Note added}---After this paper is completed, we were informed that there is a very interesting discussion on the radiated energy in Ref.~\cite{Mikhailov:2003er}. It is easy to find that
Eq.(24) of Ref.~\cite{Mikhailov:2003er} can be easily reduce to our Eq.~(\ref{re}) by the following identification. The radiation energy found in Ref.~\cite{Mikhailov:2003er} through a nonlinear wave calculation is 
\begin{equation}
E=\frac{\sqrt{\lambda}}{2\pi} \int dt \frac{\ddot{\vec{x}}^2-\left[\dot{\vec{x}}\times \ddot{\vec{x}}\right]^2}{(1-\dot{x})^3}
\end{equation}
In our special case, where the acceleration is constant and velocity is parallel to the acceleration, we find the proper acceleration can be derived from the proper velocity $\underline{u}=\frac{d}{d\tau}(x,ct)$. Thus, one finds $\underline{a}=\frac{d}{d\tau}\underline{u}=\gamma^4\ddot{x}(1,v)$ and $|a|=\gamma^3 \ddot{x}$\cite{Mukhanov}. Or one can just derive this relation from Newton's second law in relativity, 
\begin{equation}
a=\frac{F}{M}=\frac{d}{dt}\frac{\dot{x}}{\sqrt{1-\dot{x}^2}}=\gamma^3 \ddot{x}
\end{equation}
Therefore, it is straightforward to perform the trivial time integration and obtain $E=\frac{\sqrt{\lambda}}{2\pi} a^2 t$ after identifying $a$ with $\gamma^3 \ddot{x}$. This is exactly the same as we found in Eq.~(\ref{re}) . 

\emph{Acknowledgment}---I would like to thank Prof. A.H. Mueller for a lot of discussion and comments. Also I acknowledge inspiring discussion with Prof. Danial Kabat and Lam Hui, as well as Kurt Hinterbichler and Stefanos Marnerides.

\end{document}